\documentclass[twocolumn]{article}
\usepackage[hang,flushmargin,stable]{footmisc}
\usepackage{graphicx}
\usepackage{amssymb}
\usepackage{amsmath}
\usepackage{colortbl}
\usepackage{color}
\usepackage{indentfirst}
\usepackage{url}
\usepackage{subfigure}
\usepackage{multirow}
\usepackage{threeparttable}
\usepackage{tikz}
\usepackage{bbding}
\usepackage[all]{xy}
\usepackage{algorithmic}
\usepackage{array}
\usepackage{multirow}
\usepackage{flushend}
\usepackage{paralist}
\usepackage[small,bf]{caption}
\usepackage{fullpage}

\usepackage[letterpaper,hmargin=1in,vmargin=1in]{geometry}

\usepackage[bookmarks=false, colorlinks=true, plainpages = false,
  linkcolor = blue,   citecolor = blue, urlcolor = blue, filecolor = blue]{hyperref}
 
\newcommand{\descr}[1]{\bigskip \noindent \textbf{#1}}

\newboolean{removeVar}
\setboolean{removeVar}{true} 

\ifthenelse{\boolean{removeVar}}{\newcommand{\remove}[1]{}}{\newcommand{\remove}{\textcolor{green}}}

\newcommand{\Alexa}{\textsf{Alexa}}

\newcommand{\phishing}{\textsf{Phishing}}
\newcommand{\typosquatting}{\textsf{Typosquatting}}
\newcommand{\comnet}{\textsf{*.com/*.net}}

\title{\huge\bf Harvesting SSL Certificate Data to\\Identify Web-Fraud}
\author{~\\ Mishari Almishari, Emiliano De Cristofaro, Karim El Defrawy, Gene Tsudik \vspace{0.1cm}\\
 {
   University of California Irvine
   }\\
}

\date{}
\begin{document}
\maketitle
\thispagestyle{empty}
\begin{abstract}
Web-fraud is one of the most unpleasant features of today's Internet.
Two well-known examples of fraudulent activities on the web are phishing and typosquatting.
Their effects range from relatively benign (such as unwanted ads) to
downright sinister (especially, when typosquatting is combined with phishing).
This paper presents a novel technique to detect web-fraud domains that
utilize HTTPS. To this end, we conduct the first comprehensive study
of SSL certificates. We analyze certificates of legitimate and popular domains and
those used by fraudulent ones. Drawing from extensive measurements, we build
a classifier that detects such malicious domains with high accuracy. \vspace*{0.1cm}
~
\end{abstract}

 \section{Introduction}\label{sec:introduction}
The Internet and its main application -- the web -- have been growing
continuously in recent years. In the last three years, Web contents have doubled in size, 
from $10$ billion to over $20$ billion pages~\cite{wwwsize}.
Unfortunately, this growth has been accompanied by a parallel increase 
in nefarious activities, as reported by~\cite{incr}.
Indeed, the web is an appealing platform for various types of electronic fraud, such as {phishing} and
{typosquatting}.

 \emph{Phishing} aims to elicit sensitive information 
-- e.g., user names, passwords or credit card details -- from unsuspecting users. 
It typically starts with a user being directed to a fake website with
the look-and-feel of a legitimate, familiar and/or well-known one.
Consequences of phishing range from denial-of-service to
full-blown identity theft, followed by real financial losses. In 2007, more than
3 billion U.S. dollars were lost because of phishing~\cite{phishloss2007}. 

\emph{Typosquatting} is the practice of registering domain names that
are typographical errors (or minor spelling variations) of addresses of well-known
websites (target domains) \cite{Wang}. 
It is often related to domain parking services and advertisement syndication, i.e., instructing browsers 
to fetch advertisements from a server and blending them with content of the
website that the user intends to visit \cite{Wang}. In addition to displaying 
unexpected pages, typo-domains often display malicious, offensive and
unwanted content, or install malware~\cite{googkle,NDSS}. Certain typo-domains of 
children-oriented websites were found to even redirect users to adult content~\cite{microsoft}. Worse 
yet, typo-domains of financial websites can serve as natural
platforms for {\em passive} phishing attacks.\footnote{Passive phishing attacks do not rely on 
email/spam campaigns to lead people to access the fake website. Instead, users 
who mis-type a common domain name end up being directed to a phishing website.}

Recent studies have assessed popularity of both types of malicious activities, e.g., 
\cite{Sturgeon,spamscatter}.  In the fall of 2008, McAfee Alert
Labs found more than $80,000$ domains typosquatting on just the top $2,000$
websites~\cite{mcafee}. According to the Anti-phishing Working Group (APWG),
 the number of reported phishing attacks between April 2006 and April
2007 exceeded $300,000$ \cite{april07}. In  August 2009, $56,632$
unique phishing websites were reported by APWG -- the highest number in APWG's history~\cite{q3}.
Nevertheless, these problems remain far from solved. 

Our goal is to counter web-fraud by detecting domains hosting such malicious activities.
Our approach is inspired by recent discussions in the web-security community. 
Security researchers and practitioners have been increasingly advocating a transition to 
HTTPS for all web transactions, similar to that from Telnet to SSH. 
Examples of such discussions can be found in~\cite{Weaver},~\cite{Lauren}, and~\cite{httpseverywhere}. 
Our work is also a response to an alarming trend observed by a recent study~\cite{symmantec-blog}: 
the emergence of sophisticated phishing attacks abusing SSL certificates. These attacks rely on SSL 
to avoid raising users' suspicion, by masquerading as  legitimate ``secure'' websites. 

This brings us to the following questions:
\begin{enumerate}
\item To what extent is HTTPS adopted on the Internet?
\item How \emph{different} are SSL certificates used by web-fraudsters from those of legitimate domains?
\item Can we use information in SSL certificates to 
identify web-fraud activities, such as phishing and typosquatting, without compromising user privacy?
\end{enumerate}

\subsection{Roadmap}
First, we measure the overall prevalence of HTTPS in popular and randomly 
sampled Internet domains. Next, we consider popularity of HTTPS 
in the context of web-fraud by studying its use in phishing and 
typosquatting activities. Finally, we analyze, for the first time, all fields in SSL 
certificates and  
identify useful features and patterns that can help identify web-fraud.

Leveraging our measurements, we propose a novel technique to identify web-fraud domains that use
HTTPS. We construct a classifier that analyzes certificates of such domains.
We validate our classifier by training and testing it over data 
collected from the Internet. The classifier achieves a detection 
accuracy ranging from $99\%$, in the worst-case, to $96\%$. It only relies on data stored in SSL certificates and does not require any user information.
Our classifier is orthogonal to prior mitigation techniques and can be integrated with other methods 
(that do not rely on HTTPS), thus improving overall effectiveness and facilitating detection of a wider range of malicious domains.
This might encourage legitimate websites that require sensitive user information (thus, potential 
phishing targets) to enforce HTTPS.

Finally, we highligh some indirect benefits of HTTPS: it does not 
only guarantee confidentiality and authenticity, but can also help combat web-fraud.


\vspace{0.3cm}\noindent{\bf Paper Organization.} The rest of the paper is organized as follows. In Section \ref{sec:x509}, we briefly overview
X.509 certificates. Section~\ref{sec:experiments} presents the rationale and details of our measurements, as well as their analysis. 
In Section~\ref{sec:classifier}, we describe details of a novel classifier that detects malicious 
domains, based on information obtained from SSL certificates.
Section~\ref{sec:discussion} discusses implications of our findings and limitations of our solution. Section~\ref{sec:related}
overviews related work. Finally, we conclude in Section~\ref{sec:conclusion}.


\begin{figure}[t]
  \centering
  \includegraphics[width=0.35\textwidth]{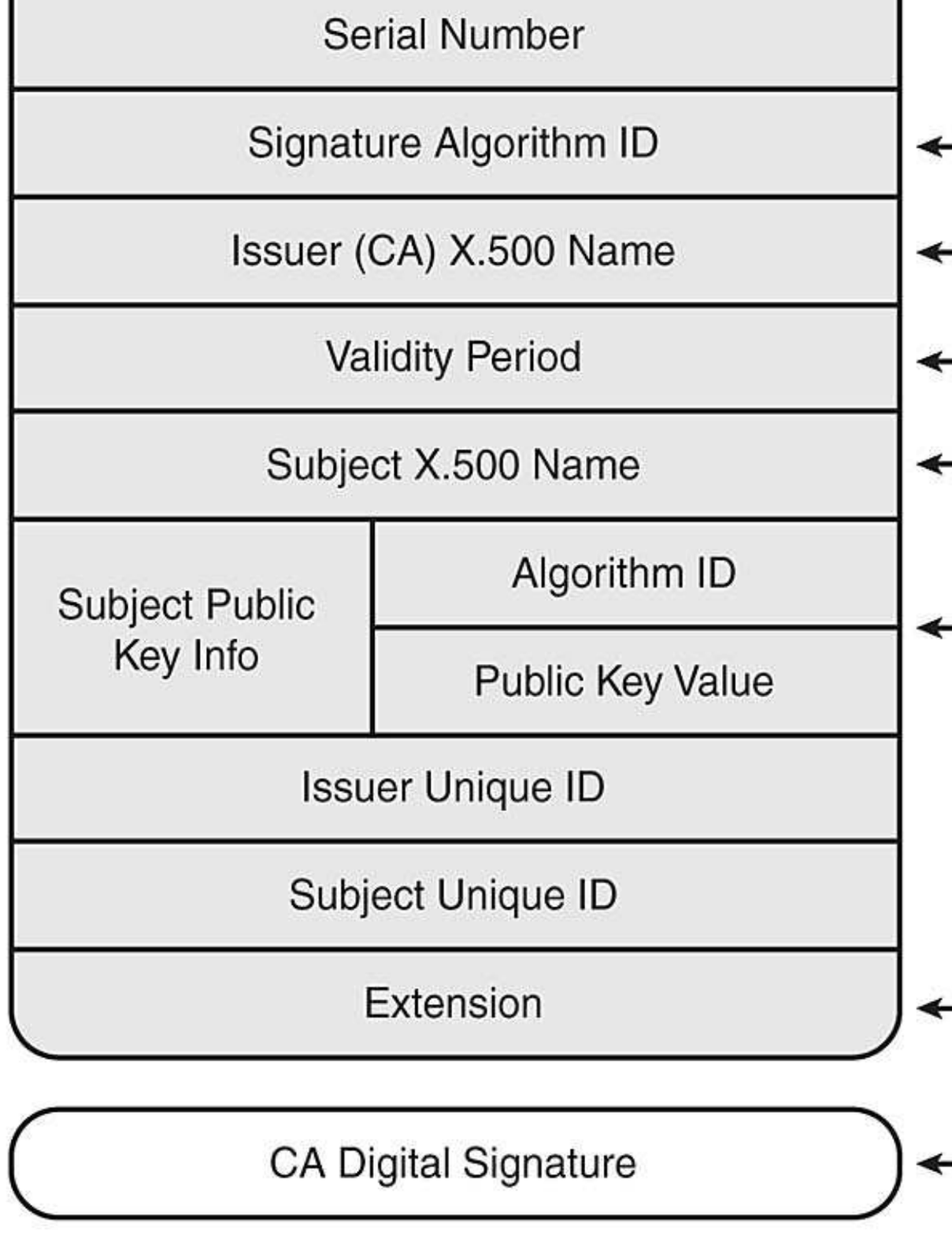}
  \caption{\label{fig:x509format} {\small \bf X.509 certificate format.\vspace{-0.4cm}}}
\end{figure}

\section{X.509 Certificates}\label{sec:x509}
The term {\em X.509 certificate} usually refers to 
an IETF's PKIX Certificate and CRL Profile of the X.509 v3 certificate standard, as 
specified in RFC 5280~\cite{x509rfc}. In this paper we are concerned with the public key 
certificate portion of X.509. In the rest of the paper, we refer to the public key certificate 
format of X.509 as a {\em certificate}, or an {\em SSL/HTTPS certificate}. 

According to X.509, a Certification Authority (CA) issues a certificate binding a public key to an X.500
Distinguished Name (or to an Alternative Name, e.g., an e-mail address or a DNS-entry). 
Web-browsers -- such as Internet Explorer, Firefox, Chrome and Safari -- come 
with pre-installed trusted root certificates. Browser software makers determine which 
CAs are trusted third parties. Root certificates can be 
manually removed or disabled, however, it is unlikely that typical users do so. 
The general structure of an X.509 v3 \cite{x509rfc} certificate is presented in Figure 
\ref{fig:x509format}. As discussed later in Section \ref{sec:experiments}, 
we analyze \textit{all fields} in certificates collected from both legitimate and malicious domains.

\section{Measurements and Analysis of SSL Certificates}\label{sec:experiments}
In this section, we describe our data sets and our collection methodology. 
We then present our analysis leading to the design of a classifier that detects web-fraud domains. 

\subsection{HTTPS Usage and Certificate Harvest} \label{subsec:method-data-sets}
Our measurement data was collected during the following periods: (1) from September to October 2009, and (2) from March to October 2010.
We include three types of domain sets: \textit{Popular, Random and Malicious (Phishing and Typosquatting)}. Each domain
was probed twice. We probed each domain for web existence (by sending an HTTP request to port 80)
and for HTTPS existence (by sending an HTTPS request to port 443). We harvested the SSL certificate when a domain responded to HTTPS. Table~\ref{table:data set-desc} lists our data sets and the number of corresponding SSL certificates.

\begin{table}[b]
\centering {\footnotesize
\begin{tabular}{|l|c|c|}
\hline
{\bf Data set} & {\bf Type } & {\bf Number of Certificates }  \\ \hline
\Alexa\ & Popular & 2,984   \\ \hline
\comnet\ & Random & 22,063   \\ \hline
\phishing\ & Malicious  & 5,175  \\ \hline
\typosquatting\ & Malicious  & 486 \\ \hline
\end{tabular}} \vspace{-0.3cm}
\caption{\label{table:data set-desc}
\textbf{\small Description of data sets and corresponding number of certificates.}}
\end{table}

\begin{table*}[t]
\centering {\footnotesize
\begin{tabular}{|l|l|l|l|}
\hline
  {\bf Feature} & {\bf Name} &{\bf Type}  & {\bf Notes} \\ \hline
F1 & md5 & boolean & The signature algorithm of the certificate is md5WithRSAEncryption \\ \hline
F2 & bogus subject & boolean  & The subject section of the certificate is clearly bogus \\ \hline
F3 & self-signed & boolean &  The certificate is self-signed \\ \hline
F4 & host-common-name-sim & boolean & Whether the common name of the subject and domain name\\
   &                      &          & of the certificate has the same basic part of the domain\\ \hline
F5 &issuer common & string & The common name of the issuer \\\hline
F6 &issuer organization & string & The organization name of the issuer \\ \hline
F7 &issuer country & string & The country name of the issuer \\ \hline
F8 &subject country & string & The country name of the subject\\ \hline
F9 & validity duration & integer & The validity period in days \\ \hline
\end{tabular}}
\vspace{-0.2cm}
\caption{\label{tbl:features} \textbf{\small Features extracted from SSL
certificates.}}
\end{table*}

\descr{Popular Domains Data Set.}
The Alexa~\cite{alexa} top $10,000$ domains were used as our data set for popular domains. 
Alexa is a subsidiary of {\sf Amazon.com} known for its
web-browser toolbar and for reporting traffic ranking of Internet websites.
The company does not provide exact information on the number
of users using its toolbar, but it claims several millions \cite{alexa-ranking}.
We use Alexa ranking as a popularity measure for legitimate domains. From the top $10,000$ \Alexa\ domains, we collected 
certificates from $2,984$ different domains (Table \ref{table:data set-desc}). Only $2,679$ of these certificates are unique. $13.5\%$ of domains have identical certificates to some other domain in the same data set. Some popular domains belong to the same owner and represent the same company, e.g., \url{google.co.in} and \url{google.de}. In this paper, we refer to this data set as \Alexa\ set. 

\descr{Random Domain Data Set.}
We randomly sampled .com and .net domains to create a data set that represents a random portion of 
the web, i.e., allegedly benign but ``unpopular'' domains. We 
created this data set to test whether (or not) certificates of random domains contain similar features to those of either popular 
or malicious domains. We randomly sampled $100,000$ domains from the .com Internet Zone File (and $100,000$ from the .net counterpart).
The .com/.net Internet Zone Files have been collected from VeriSign \cite{verisignurl}. After sampling the Zone Files, we downloaded 
existing SSL certificates and, as shown in Table \ref{table:data set-desc}, we obtained $22,063$ different domains 
with their SSL certificates. Some of these certificates were duplicates($16,342$ unique certificates). The 
percentage of domains having the same certificate as another domain in the same data set in this case is $36.4\%$. In this paper, we 
refer to this data set as \comnet\ set.

\subsubsection{Malicious Data Set}
\noindent{\bf Phishing.} We collected SSL certificates of $5,175$ different domains of phishing urls. The number of 
unique certificates is $2,310$ (Table \ref{table:data set-desc}). Phishing domains were obtained from the PhishTank 
website \cite{phishtank} as soon as they were listed. Reported
URLs in PhishTank are verified by several users and, as a result, are malicious with very high
probability. We consider this data set as a baseline for phishing domains. The percentage of domains that have an identical certificate as another 
domain within the data set is $63\%$. This is a significant increase compared to the popular and random sets. One possible explanation 
is related to the fact that a large portion of phishing certificates are self signed. These 
certificates are generated locally and then re-used for several domains under the control of the same 
malicious entity. In this paper, we refer to this data set as \phishing\ set. 

\descr{Typosquatting.} 
In order to collect SSL certificates of typosquatting domains,
we first identified the typo domains in our \comnet\ random domains  by 
using Google's typo correction service~\cite{GoogleSpellChecker}. 
This resulted in $38,617$ possible typo domains. 
However, these might be benign domains that {\em accidentally} resembled typos of well-known domains. 
We identified the typosquatting domains in this set by detecting the parked domains 
among them, using the machine-learning-based  classifier proposed in~\cite{ads-portal-class}.
Note that typosquatters profit 
from traffic that accidentally comes to their domains. 
One common way of achieving this is to host typosquatting domains from a domain parking 
service. Parked domains~\cite{parked} are ads-portal domains that 
show ads provided by a third-party service called a parking service, in the 
form of ads-listing~\cite{Wang} so typosquatters may profit from incoming traffic 
(e.g., if a visitor clicks on a sponsored link). 
We discovered that $9,830$ out of $38,617$ were parked domains. We consider these $9,830$ names 
as the data set of typosquatting domains. We then probed these domains to get their SSL certificates. 

As reported in Table \ref{table:data set-desc}, our \typosquatting\ data set is composed of $486$ domains, i.e., the parked domains 
having HTTPS and responding with SSL certificates.  However, note that only $100$ out of $486$ certificates are 
unique. The percentage of domains that have a duplicate certificate with another
domain in this data set is $87\%$. In this paper, we refer to this data set as \typosquatting\ set.

We acknowledge that the size of the \typosquatting\ data set is relatively limited. Therefore, we do not claim conclusive results
from its analysis, but we include it for completeness. We believe that the limited size of the \typosquatting\ set is due to the 
lack of incentives from using HTTPS in this context. Using HTTPS in typosquatting domains does not help in luring users 
(unlike \phishing). Nonetheless, we believe that, being the first of its kind, such an analysis of HTTPS-enabled typosquatting 
domains is interesting and shows an initial insight into this fraudulent activity. 

\subsection{Certificate Analysis}\label{analysis-ssl}
The goal of this analysis is to guide the design of our detection method, i.e., the classifier presented in Section~\ref{sec:classifier}.
One side-benefit is to reveal differences
between certificates used by fraudulent and legitimate/popular domains.
In total, we identified $9$ relevant certificate features, listed in Table
\ref{tbl:features}. (Other features and fields (e.g.,
RSA exponent, Public Key Size, ...) had no substantial differences
between legitimate and malicious domains and we omit them here.)
Most features map to actual certificate fields, e.g., F1 and F2. Others are
computed from certificate fields but are not
directly reflected in the certificate, e.g., host-common-name-sim (F4).
Some features are boolean, whereas,
others are integers, reals or strings. The computation of features is performed on 
all the certificates (including the duplicates) as our goal is to identify malicious domains (regardless of its certificates being duplicates or not).

\subsubsection{Analysis of Certificate Boolean Features}
%
Features F1-F4 have boolean values, e.g., F1 (md5) is true if the signature algorithm used 
in the certificate is ``md5WithRSAEncryption.'' The results of analyzing these features 
are summarized in Table \ref{tbl:boolean-features-analysis}. This analysis reveals
interesting and unexpected issues and differences between legitimate and malicious domains.

\begin{table}[t]
\centering {\footnotesize
\begin{tabular}{| l | l | l | l | l | }
\hline
  {\bf Feature}\hspace{-0.1cm} & {\bf \Alexa\ \hspace{-0.3cm}} & {\bf \comnet\ \hspace{-0.2cm}} & {\bf \phishing\ \hspace{-0.2cm}} & {\bf \typosquatting\ \hspace{-0.2cm}} \\\hline
F1 & $9.9\%$ & $27.4\%$& $17.4\%$ & $26.1\%$\\ \hline
F2 & $7.7\%$ & $11.8\%$& $18.3\%$ & $29.8\%$\\ \hline
F3 & $15.8\%$& $35.4\%$& $30.6\%$ & $53.5\%$\\ \hline
F4 & $74.3\%$& $10.4\%$& $9.8\%$ & $0\%$\\ \hline
\end{tabular}}
\vspace{-0.2cm}
\caption{\label{tbl:boolean-features-analysis} \textbf{\small Analysis of boolean certificate features
(percentages satisfying each feature).}}
\end{table}

\subsubsection*{F1 (md5)} \noindent 
Observe that $9.9\%$ of \Alexa\ certificates use ``md5WithRSAEncryption'', much less than those in \comnet\ ($27.4\%$), 
\phishing\ (17.4\%) and \typosquatting\ (26.1\%). 
Note that rogue certificates can be constructed using MD5 (\cite{stevens2007chosen}, \cite{stevens2009short}). 
The difference in the percentages may indicate that  many of \comnet, \phishing\ and \typosquatting\ certificates are issued without a
significant effort to check their security.

\subsubsection*{F2 (bogus subject)} \noindent 
A bogus subject indicates whether the subject fields have some
bogus values (e.g., ``ST=somestate'',  ``O=someorganization'',
``CN=localhost'', ...). We identified a list of such meaningless values and 
considered subjects to be {\em bogus} if they contain one of these values. 7.7\% of \Alexa\ certificates
satisfy this feature (similar percentage in \comnet\ ). This percentage is much higher (18.3\%) in \phishing\ and \typosquatting\ (29.8\%) .
This indicates that web-fraudsters fill subject values
with bogus data or leave default values when generating certificates.

\subsubsection*{F3 (self-signed)} \noindent 
15.8\% of \Alexa\
certificates are self-signed, somehow unexpectedly. One possible explanation is that some of the popular domains in 
the  \Alexa\ set represent companies having their own CA and issuing their own certificates (e.g., google, microsoft, yahoo ...etc.).
The percentages of self-signed certificates in \comnet, \phishing, and \typosquatting\ is higher (resp.,  35.4\%, 30.6\% and 53.5\%). 
This is expected for \phishing, since miscreants would like to avoid leaving any trace by obtaining a certificate from a 
CA (which requires documentation and a payment). For \comnet, a higher percentage could be explained  by the use of locally generated certificates, e.g.,  using OpenSSL \cite{openSSL}.
The difference between \Alexa\ on one side and \phishing, \typosquatting, and \comnet\ on the other side is quite significant (more than 14\%). 

\subsubsection*{F4 (host-common-name-sim)}\noindent 
We expect the common name in the subject field to be very similar to the hostname in popular domain certificates, while
we intuitively expect this to be lower in malicious domain certificates, e.g., because malicious domains may not use
complying SSL certificates. In order to assess this, we define a feature called ``host-common-name sim``. This feature 
measures the similarity between domain name of the SSL certificate and common name of the subject field in it.
For instance, if the hostname and common name are \texttt{www.google.com.sa} and \texttt{google.com} respectively, 
host-common-name-sim is set to true. Whereas, if the hostname and common name are equal 
to\texttt{ www.domain-x.com} and \texttt{www.domain-y.com}, respectively, the feature is set to false since domain-x is not equal to domain-y.

74.3\% of \Alexa\
certificates satisfies this feature. The percentages in  \comnet, \phishing, and \typosquatting\ are 10.4\% and 9.8\%, 0\%
respectively. The difference between \Alexa\ on one side and \phishing, \typosquatting, and \comnet\ on the other side is quite remarkable. 
This feature, together with the previous one (excluding bogus subject), suggests that there is some strong similarity among 
\comnet, \phishing, and \typosquatting\ certificates. 

\subsubsection{Analysis of Certificate Non-Boolean Features}
\label{non-boolean-analysis}
We now present the analysis of non-boolean features. F5--F6 are related to the certificate issuer:
common name, organization name and country, while F7--F8 are related to the country issuer or country subject and F9 is related the validity duration.

\begin{table}[t]
\centering {\footnotesize
\begin{tabular}{| l | l | l | }
\hline
  {\bf Feature} & {\bf Sets } & {\bf APCR  }  \\\hline
issuer common name  & \phishing\,\Alexa\ & 52.7\%\\ \hline
issuer common name  & \phishing\,\comnet\ & 51.2\%\\ \hline
issuer organization name  & \phishing\,\Alexa\ & 57.7\%\\ \hline
issuer organization name  & \phishing\,\comnet\ & 46\%\\ \hline
issuer country  & \phishing\,\Alexa\ & 28.4\%\\ \hline
issuer country  & \phishing\,\comnet\ & 18.6\%\\ \hline
subject country & \phishing\,\Alexa\ & 26.5\%\\ \hline
subject country & \phishing\,\comnet\ & 24\%\\ \hline
\end{tabular}}
\vspace{-0.2cm}
\caption{\label{tbl:apcr-values} \textbf{\small APCR values.}}
\end{table}

\begin{table*}[t]
\centering {\footnotesize
\begin{tabular}{| l | l | l | l |}
\hline
  {\bf Feature} & {\bf Sets } & {\bf Ratio\_A }  & {\bf Ratio\_B}\\\hline
issuer common name  & \phishing\,\Alexa\ & 72\% & 23\%\\ \hline
issuer common name  & \phishing\,\comnet\ & 79\% & 31\%\\ \hline
issuer organization name  & \phishing\,\Alexa\ & 88\% & 33\%\\ \hline
issuer organization name  & \phishing\,\comnet\ & 76\% & 34\%\\ \hline
issuer country  & \phishing\,\Alexa\ & 92\% & 65\%\\ \hline
issuer country  & \phishing\,\comnet\ & 33\% & 15\%\\ \hline
subject country & \phishing\,\Alexa\ & 55\% & 30\%\\ \hline
subject country & \phishing\,\comnet\ & 60\% & 28\%\\ \hline
\end{tabular}}
\vspace{-0.2cm}
\caption{\label{tbl:phish-alexa-com} \textbf{\small The table shows the ratios of Popular-Issuer-Common-Name-A-B, Popular-Issuer-Organization-A-B, Popular-Issuer-Country-A-B and Popular-Subject-Country-A-B in set A and B. In this table, A is \phishing\ and B is either \Alexa\ or \comnet. }}
\end{table*}

\subsubsection*{F5 (issuer common name) and F6 (issuer organization)}\noindent
First, we noticed that some values of issuers' common names are popular in only one domain set. For example, 10.1\% of  \phishing\ certificates are 
issued by UTN-USERFIRST-HARDWARE (only 4.2(5.3)\% in \Alexa(\comnet)). Similarly, some issuers'  organization names are popular only in one domain set. For example, 9.6\% of  \phishing\ certificates have COMODO-CA-LIMITED
as their issuer's organization name, as opposed to 2.1\% in \Alexa\ and 3.2\% in \comnet\ certificates. 

To elaborate more on the difference of the issuer common name between any two domain data sets, \textbf{A} against  \textbf{B}, we extract all issuer common names that are more popular(in terms of ratio) in set \textbf{A} than in \textbf{B}. We name these popular common names \textbf{Popular-Issuer-Common-Name-A-B}. Then, we compute what percentage of the certificates in \textbf{A} and \textbf{B} that have their issuer common name in \textbf{Popular-Common-Name-A-B}. In a similar way, we extract \textbf{Popular-Issuer-Organization-A-B}, \textbf{Popular-Issuer-Country-A-B}, and \textbf{Popular-Subject-Country-A-B}  sets corresponding to issuer organization, issuer country, and subject country. Table \ref{tbl:phish-alexa-com} shows the ratios when \textbf{A} is \phishing\ and \textbf{B} is \Alexa\ or \comnet. For issuer common name(issuer organization name), we can clearly notice the difference in ratios between \phishing\ and \Alexa(\comnet)  which emphasizes on the differences of issuer common name(issuer organization name) feature among the domain sets (The ratios when \textbf{A} is \Alexa\ or \comnet\ suggest similar conclusions). 

To quantify the difference in issuer's common/organization name, we measure change in the posterior probabilities of phishing certificates, i.e., the probability that a certificate is phishing given the common/organization name is equal to a specific value. To this end, we merge \phishing\ and \Alexa\ sets and observe how the posterior probability changes. Similarly, we merge \phishing\ and \comnet\ sets. In order to measure the changes in the posterior probability, we borrow the metric in \cite{ads-portal-class}, i.e., called Average Posterior Change Ratio $APCR$. In the following we define $APCR$. Note that $P_p$ stands for posterior probability of phishing certificates and $P_a$ stands for prior probability of being a phishing certificate. In the following, $I$ stands for a common name issuer value\footnote{The $APCR$ defined for issuer's common name  can be similarly defined for issuer's organization name, issuer's country, and subject country.}:

{\footnotesize
$$APCR=\sum_{\forall I}P_p \mbox{ Change Ratio in } I \times \mbox{Fraction of Certif. in } I$$ }
where
{\footnotesize
$$P_p\ Change\ Ratio\ in\ \textit{I}\ =\ \frac{\lvert P_p\ in\ \textit{I} - P_a \rvert}{Max\ P_P\ Displacement\ in\ \textit{I}}$$
}
and
{\footnotesize
$$Max\ P_p\ Displacement\ in\ \textit{I} = \left\{ \begin{array}{rl}
  1\ -\ P_a &\mbox{if} P_p in\ \textit{I}\ >\ P_a \\
  P_a &\mbox{ otherwise}\end{array}\right.$$
 }


%
%
The Posterior Change Ratio measures relative change of posterior probability $P_p$ from the prior probability
, $P_a$. For instance, a value of 0 indicates no relative change and a value of 
1 indicates the largest relative change. Note that 
the average is weighted, such that more ``weight'' is given to common names 
corresponding to more certificates  (For more information, readers can refer to \cite{ads-portal-class}). 

Table \ref{tbl:apcr-values} shows the $APCR$ of for issuer's common name. When we merge \phishing\ and \Alexa\ sets, we obtain an $APCR$ value 
of 52.7\% (51.2\% for \phishing\ and \comnet). This indicates that change in posterior probability is quite significant.
Table \ref{tbl:apcr-values} shows similar  $APCR$ values for issuer's organization name.

\subsubsection*{F6 and F7 (issuer and subject countries)} \noindent
Similarly, some of issuers' country names are popular only in  one data set. For example, 10.3\% of  \phishing\ certificates have {\tt GB} as their issuer's country name
(only 1.4\% in \Alexa\ and 4.3\% in \comnet). Additionally, some of the subject country names are popular only in the \phishing\ set. For example, 5.3\% of  \phishing\ certificates have FR  as their country in the subject field. This happens only in 2.2\% of \Alexa\ certificates (0.9\% in \comnet).

Table \ref{tbl:phish-alexa-com}, shows the ratios of domains that have their certificates issuer country(subject country) in \textbf{Popular-Issuer-Country-A-B} (\textbf{Popular-Subject-Country-A-B}) set where \textbf{A} is \phishing\ and \textbf{B} is \Alexa\ or \comnet. The ratios in \phishing\ and \Alexa(\comnet) have significant differences which emphasizes on the differences of issuer country(subject country) feature among the domain sets. Table \ref{tbl:apcr-values} shows the $APCR$ of the issuer country  feature. When we combine the \phishing\ and \Alexa\ sets, we obtain an $APCR$ value of 28.4\% (18.6\% for the \phishing\ and \comnet\ sets). This shows that there is a change in the posterior probability, however, not as significant as for the issuer common/organization name. Table \ref{tbl:apcr-values} also shows similar  $APCR$ values for subject country feature which suggest similar conclusions.

\subsubsection*{F9 (validity duration)} 
\begin{figure}[t]
  \small
  \centering
   \includegraphics[width=0.48\textwidth]{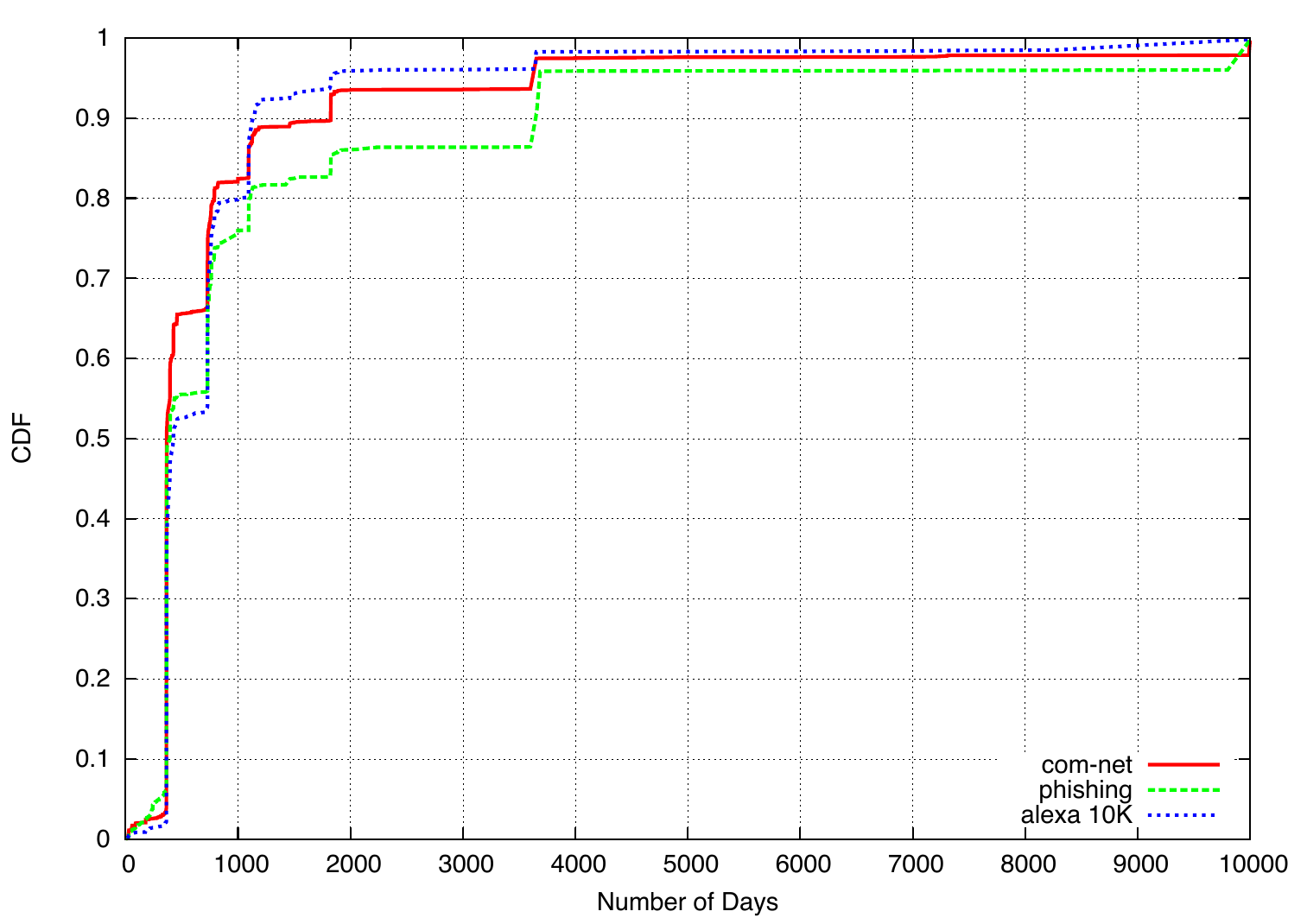}
   \vspace{-0.4cm}
   \caption{\label{fig:ccdf-cert-duration} {\small \bf CDF of certificate duration.}}
\end{figure}

\noindent
Table \ref{tbl:duration}, shows the distribution of domains over different duration periods. As shown in the Table, domains sets have different ratios for different periods. For example, \phishing\ has the largest ratio when the period is less than a year, \comnet\  when the period is in [365-729] days, and \Alexa\ when the period is in [730-1094] days. Additionally, we plot the CDF of the certificate duration for different data sets in Figure \ref{fig:ccdf-cert-duration} which elaborates more on the differences.  

\begin{table}[t]
\centering {\footnotesize
\begin{tabular}{| l | l | l | l | l | }
\hline
  {\bf Feature} & {\bf \Alexa\ } & {\bf \comnet\ } & {\bf \phishing\ } \\\hline
0 - 364 & $2.5\%$ & $4.2\%$& $6.3\%$ \\ \hline
365 - 729 & $51.7\%$ & $62.3\%$& $49.6\%$  \\ \hline
730 - 1094 & $26.1\%$& $16.1\%$& $20.1\%$  \\ \hline
1095 - 3649 & $15.9\%$& $11.1\%$& $10.4\%$ \\ \hline
3649 -  & $3.8\%$& $6.3\%$& $13.5\%$ \\ \hline
\end{tabular}}
\vspace{-0.1cm}
\caption{\label{tbl:duration} \textbf{\small The table shows the distribution of domains over different duration periods for different domain sets.}}
\end{table}

\subsubsection{Certificate Feature Analysis}
We now discuss the most important observations:
\begin{enumerate}
\item For most of the features, distributions of malicious certificates are significantly different from \Alexa\ certificates. 
Therefore, popular domains can be easily differentiated from malicious domains based on their SSL certificates.

\item A self signed certificate is more likely to be for popular domain. The percentage of self signed certificates in popular domains is only 15.8\%.

\item We observe strong similarities between the \comnet\ set and malicious sets in many features. One reason is that certificates in both sets 
may be issued without applying appropriate control, as opposed to certificates obtained from popular \Alexa\ domains. Another possible 
explanation is that a large portion of  certificates from \phishing\, \typosquatting\ and \comnet\ domains are locally generated 
(e.g., self-signed) and use the same default values
from the employed software (e.g., OpenSSL).

\item Host-common-name-sim feature is a very discriminative feature in differentiating between popular and other domains. 
  
\end{enumerate}

\section{Certificate-based Classifier}\label{sec:classifier}
The analysis above shows that several features have  
distributions that vary among different data sets. Relying on a single feature to identify malicious certificates will yield a high rate of false positives. 
Therefore, we combine all features and feed them to a set of machine learning classifiers in order to differentiate among certificates belonging to different data sets. We use several machine-learning-based classification algorithms and select the one with the best performance.
Specifically, we consider the following algorithms: Random Forest \cite{randforest}, Decision Tree \cite{dtree}, and  Nearest Neighbor \cite{mlearn1}. In addition, we explore two optimization techniques for
Decision Trees: Bagging \cite{bagging, boost-bag-c4.5} and Boosting \cite{boosting, boost-bag-c4.5}.

Nearest Neighbor classifier does not have a training phase and it simply labels the incoming test record to the most similar training record. Decition Tree classifier builds a tree that classifies a record by following a path from the root to the leaf. To select a specific branch, the record has to satisfy its condition. The Random Forest classifier is derived from a multiple of Decision Trees each of which has some selection of the feature set. Since, detailed algorithm descriptions are out of the scope of this paper,
we refer to \cite{mlearn1,bagging,boost-bag-c4.5,boosting,boost-bag-c4.5}
for relevant background information. 

We use precision-recall performance metrics to summarize performance of a classifier. These metrics consists of the following
ratios: Positive Recall, Positive Precision, Negative Recall, and Negative Precision. We use the term ``Positive'' set to
denote \phishing\ (and  \comnet\ in one case) certificates  and ``Negative'' set to refer to \Alexa\ (popular) certificates.  The positive
(negative) recall value is the fraction of positive (negative) domains that are correctly identified by the classifier.
Positive (negative) precision is the actual fraction of positive (negative) domains in the set identified by the classifier.

We evaluate the performance of the classifier using the ten-fold cross validation method \cite{mlearn1}.
We randomly divide the data set into 10 sets of equal size, perform 10 different training/testing steps
where each step consists of training a classifier on 9 sets, and then we test it on the remaining set.
Finally, we average all results to assess the final performance of the classifier.

\subsection{Classifier Features}\label{sec:classfeatures}
For feature F5 - issuer common name, we extract 6 boolean sub-features each of which corresponds to whether the domain certificate issuer common name belongs to \textbf{Popular-Issuer-Country-A-B}  for a specific pair selection (\textbf{A}, \textbf{B}). We try all the possible pair combinations; specifically, (\phishing, \Alexa), (\phishing, \comnet), (\Alexa,\comnet), (\Alexa,\phishing), (\comnet,\Alexa), and (\comnet,\phishing). This gives us the boolean sub-features \textbf{issuer-common-name-A-B}. Similarly, we extract 6 sub-features from F6, F7, and F8 giving us \textbf{issuer-organization-A-B}, \textbf{issuer-country-A-B}, and \textbf{subject-country-A-B} sub-features for 6 different pairs of ($A$,$B$). It is these sub-features that we use in the classifiers instead of F5-F8. 

For feature F9 - validity duration, we convert it to a nominal feature before feeding it to the classifier. That is, for each domain validity duration, we convert it to a nominal value by assigning it one of five values corresponding to different duration periods shown in Table \ref{tbl:duration}. 

For the other features, we use them as they are. Note that in the classifier the Negative set is \Alexa\ and the positive set could be either \phishing\ or \phishing\ and \comnet. When the positive set is both \comnet\ and \phishing\ we use F1-F4, sub-features of F5-F8, and the converted version of F9. When the positive set is only \phishing, we use the same features but we exclude all the sub-features that has \comnet\ in the pair combination.

\subsection{Classifier Results}\label{sec:classresults}
\begin{table}[t]
\centering
{\footnotesize
\begin{tabular}{|c|c|c|c|c|}
\hline

Classifier & Positive & Positive & Negative & Negative\\
        & Recall & Precision & Recall & Precision\\
\hline
Random & 0.94 & 0.88 & 0.778  & 0.881\\
Forest &&&&\\
\hline
Decision Tree  & 0.939 & 0.881  & 0.779  & 0.88\\
\hline
Bagging & 0.935 & 0.877 & 0.773  & 0.874 \\
Decision Tree  & & &  & \\
\hline
Boosting  & 0.94 & 0.881 & 0.78  & 0.882\\
Decision Tree  & & &  & \\
\hline
Nearest  & 0.94  & 0.879  & 0.774  & 0.882 \\
Neighbor&&&&\\
\hline
\end{tabular}
}
\vspace{-0.1cm}
\caption{\bf Performance of classifiers - data set consists of: (a)positive: \phishing\ certificates
 and (b) negative: \Alexa\ certificates.}
\label{tab:class-1}
\end{table}

\begin{table}[t]
\centering
{\footnotesize
\begin{tabular}{|c|c|c|c|c|}
\hline

Classifier & Positive & Positive & Negative & Negative\\
        & Recall & Precision & Recall & Precision\\
\hline
Random & 0.974 & 0.958 & 0.61  & 0.72\\
Forest &&&&\\
\hline
Decision Tree  & 0.975 &  0.958 & 0.611  & 0.727\\
\hline
Bagging - & 0.972 & 0.96 & 0.631  & 0.708 \\
Decision Tree  & & &  & \\
\hline
Boosting - & 0.97.4 & 0.95.7 & 0.604   &0.719 \\
Decision Tree  & & &  & \\
\hline
Nearest  & 0.975  & 0.957  & 0.598  & 0.725 \\
Neighbor&&&&\\
\hline
\end{tabular}
}
\vspace{-0.1cm}
\caption{\bf Performance of Classifiers - data set consists of: (a)positive: \comnet\ and \phishing\  certificates
 and (b) negative: \Alexa\  certificates.}
\label{tab:class-2}
\end{table}

We first train the classifier on a data set that consists of \Alexa\ and \phishing\ certificates. The purpose is to train the
classifier to differentiate malicious from popular certificates. Performance results of different
classifiers are shown in Table \ref{tab:class-1}. Most of the classifiers have comparable performance results and the phishing detection accuracy could be as high as 88\%.

Since \comnet\ certificates have similar distributions as \phishing\ certificates in a number of features, we also build a classifier that differentiates between 
very popular certificates (\Alexa\ set) and non popular sets (\comnet\ and \phishing\ certificates). The Positive data set consists of all 
certificates from \comnet\ and \phishing\ sets. The Negative data set consists of all certificates in the \Alexa\ set. 
One can regard a certificate in the Positive set as a certificate issued with due diligence, unlike
one in the Negative set. Thus, this classifier differentiates ``neat'' certificates from ``sloppy'' ones, indicating that the corresponding
domain might be malicious. The results of our training is shown in Table \ref{tab:class-2}. Note that all classifiers have relatively similar results and the malicious detection accuracy could be as high as 96\%. 

Similar to machine-learning-based spam filtering solutions, a larger training data set results in better performance.
We acknowledge that our solution would incur false positives when actually deployed. However, the number of false positives
can be reduced by training on larger data sets and constantly updating training samples (See Section \ref{setsizes}).

The best performing classifiers in both cases can be used to label domains by classifying their certificates (labels are: phishing and legitimate in the first classifier, and suspicious and non-suspicious in the second classifier.).  The classifiers work by first extracting the features from the SSL certificates and then feeding them to the classifier model which gives the labels as a result. We envision our solution as a part of larger phishing mitigation system and the role of our solution is just to label the domains with what it thinks about the domains certificates. Thus, the results of our classifiers are served as inputs to a larger phishing mitigation system that would combine our recommendations with other observations to provide an overall more accurate judgment. This larger phishing mitigation system can be either at the client machines doing the mitigation as the user browses the Internet or  at the server machines taking a preemptive investigation of some suspicious domains to either blacklist them or block them in case they turn out to be phishing.

\subsection{Classifier Results with Different Set Sizes}\label{setsizes}
To verify how the data set size would affect the classifier performance, we use as the negative set the same \Alexa\ we use in the previous section. But for the positive set, we use 3 \phishing\ sets of different sizes; namely, 500, 1000, and 2000. The ten-fold cross validation results of Decision Tree is shown in Table \ref{tab:class-3}. We can see as we increase the size of the positive set, positive precision and recall increase. Thus, a large size of \phishing\ set is essential in having a high \phishing\ detection accuracy. 

\begin{table}[t]
\centering
{\footnotesize
\begin{tabular}{|c|c|c|c|c|}
\hline

Classifier & Positive & Positive & Negative & Negative\\
        & Recall & Precision & Recall & Precision\\
\hline
Decision  & 0.732 &  0.819 & 0.973  & 0.956\\
Tree (500) &&&&\\
\hline
Decision  & 0.817  & 0.827  & 0.943  & 0.939 \\
Tree (1000) &&&&\\
\hline
Decision  & 0.826  & 0.875  & 0.921  & 0.888 \\
Tree (2000) &&&&\\
\hline
\end{tabular}
}
\vspace{-0.1cm}
\caption{\bf Performance of Classifiers - data set consists of: (a)positive: \phishing\  certificates (500, 1000, 2000)
 and (b) negative: \Alexa\  certificates.\vspace{0.5cm} }
\label{tab:class-3}
\end{table}

\subsection{Classifier Results With Minimal Set of Features}\label{minsetfeatures}
To see how the classifier perform when we use a feature set that is less vulnerable to being manipulated, we restrict the feature set to only the sub-features related to the issuer; namely, issuer common name, issuer organization and issuer country. We use \Alexa\ as our negative set and \phishing\ as our positive set. Table \ref{tab:class-4} shows the results of the  Decision Tree (the others are of comparable performance) when we only use the issuer related sub-features. Even though we use less features, we still get reasonably good positive precision and recall. 

\begin{table}[t]
\centering
{\footnotesize
\begin{tabular}{|c|c|c|c|c|}
\hline

Classifier & Positive & Positive & Negative & Negative\\
        & Recall & Precision & Recall & Precision\\
\hline
Decision  & 0.895 &  0.823 & 0.667  & 0.785\\
Tree&&&&\\
\hline
\end{tabular}
}
\caption{\bf Performance of Classifiers - data set consists of: (a)positive: \phishing\  certificates
 and (b) negative: \Alexa\  certificates. }
\label{tab:class-4}
\end{table}

\section{Discussion}\label{sec:discussion}
Based on measurements presented in previous sections, 
we find that a significant percentage of well-known domains
\textit{already} use HTTPS. It is possible to harvest their certificates
for our classification purpose, without requiring any modifications on the domains' side.
Furthermore, the non-trivial portion of phishing websites utilizing HTTPS highlights the need to analyze 
and correlate information provided in their certificates.

\descr{Using information in certificates.}
Our results show significant differences between certificates
of popular domains and those of malicious domains. Not only is this information alone 
sufficient to detect fraudulent activities as we have shown, but it is also a useful 
component in assessing a website's degree of trustworthiness, thus improving prior metrics, such
as~\cite{wu2006,protectphish,learnphishemails}. Our method should be integrated with other techniques 
to improve the effectiveness of detecting malicious domains.

\descr{Keeping state of encountered certificates.}
We deliberately chose to conduct our measurements as general as possible, without relying on user
navigation history or on user specific training data. These components are fundamental
for most current mitigation techniques~\cite{learnphishemails,phoolproof}.
Moreover, we believe that  keeping track of navigation history is detrimental to user
privacy. However, our work yields effective detection by analyzing certain coarse-grained information extracted
from server certificates and not specific to a user's navigation patterns.
This does not violate user privacy as keeping fine-grained navigation history would.

\descr{Limitations.} 
We acknowledge that additional data sets of legitimate domains need to be
taken into consideration, e.g. popular websites from DNS logs in different organizations and countries.
Data sets of typosquatting domains can be strengthened by additional and more
effective name variations. 
Also, we acknowledge that our phishing classifier may incur false positives when actually 
deployed. However, this is a common problem to many machine-learning-based mitigation solutions 
(e.g., spam filtering and intrusion detection based on machine-learning techniques) and the number of false 
positives can be minimized by training the classifier on larger and more comprehensive data 
sets. Our classifier does not provide a complete 
standalone solution to the phishing threat since many domains do not have HTTPS.
Instead, integrated with pre-existing solutions (e.g.,~\cite{wu2006,protectphish,learnphishemails}),
it improves their effectiveness in the context of 
potentially critical applications enforcing HTTPS. \\

\descr{How malicious domains will adapt.} 
Web-fraudsters are diligent and quickly adapt to new security
mechanisms and studies that threaten their business. We hope that this work will raise 
the bar and make it more difficult for web-fraudsters to deceive users.  
If web-browsers use our
classifier to analyze SSL certificate fields, we expect
one of two responses from web-fraudsters:
(1) to acquire legitimate certificates from CAs and leave a paper trail pointing to
"some person or organization" which is connected to such malicious activities, (2) to craft
certificates that have similar values to those that are most common in certificates 
of legitimate domains. Some fields/features will be easy to forge with legitimate values
(e.g., country of issuer, country of subject, subject common and organization name, validity period,
signature algorithm, serial number ...etc). For some other fields this will not be possible
(issuer name, signature ...etc) because otherwise the verification of
the certificate will fail. In either case the effectiveness of web-fraud will be reduced. Additionally, we show in Section \ref{minsetfeatures} how the classifier still performs well when only relying on issuer related features. 

\section{Related Work}\label{sec:related}
The work in \cite{ssl-crypt-strength} conducted a study to measure the 
cryptographic strength of more than 19,000 public servers running SSL/TLS. The study reported that
 many SSL sites still supported the weak SSL 2.0 protocol. Also, most of the probed servers
 supported DES, which is vulnerable to exhaustive search. Some of the sites used RSA-based 
authentication with only 512-bit key size, which is insufficient. Nevertheless, it
showed encouraging measurement results, e.g., the use of AES as default option for most of 
the servers that did not support AES. Also, a trend toward using a stronger cryptographic function 
has been observed over two years of probing, despite a slow improvement. In \cite{ssl-priv-public}, 
the authors performed a six-month measurement study on the aftermath of the discovered vulnerability 
in OpenSSL in May 2008, in order to measure how fast the hosts recovered from the bug and changed 
their weak keys into strong ones. 
Through the probing of thousands of servers, they showed that the replacement of the weak keys was 
slow. Also, the speed of recovery was shown to be correlated to different SSL certificate characteristics 
such as: CA type, expiry time, and key size. The article in \cite{ssl-survey-2000} presented a profiling 
study of a set of SSL certificates.  Probing around  9,754  SSL servers and collecting 8,081 SSL certificates, 
it found that  around 30\% of responding servers had weak security (small key size, supporting only SSL 2.0,...), 
10\% of them were already expired and 3\% were  self-signed. Netcraft \cite{netcraft-survey} conducts a monthly 
survey to measure the certificate validity of Internet servers. Recently, the study showed that 25\% of the 
sites had their certificates self-signed and less than half had their certificates signed by valid CA. Symantec 
\cite{symmantec-blog} has observed an increase in the number of URLs abusing SSL certificates. Only in the 
period between May and June 2009, 26\% of all the SSL certificate attacks have been performed 
by fraudulent domains using SSL certificates. Although our measurement study conducts a profiling of 
SSL certificates, our purpose is different from the ones above. We analyze the certificates to show how 
malicious certificates are different from benign ones and to leverage this difference in designing a mitigation technique.  

The importance and danger of web-fraud (such as phishing and typosquatting)
has been recognized in numerous prior publications, studies and industry 
reports mainly due to the tremendous financial losses 
\cite{phishloss2007} that it causes. One notable study is \cite{spamscatter} which analyzed the
infrastructure used for hosting and supporting Internet scams, including
phishing. It used an opportunistic measurement technique that 
mined email messages in real-time, followed the embedded
link structure, and automatically clustered destination websites using image
shingling. In \cite{learnphishemails},  a machine learning based methodology was 
proposed for detecting phishing emails. The methodology was based on a classifier 
that detected phishing with 96\% accuracy and false negative rate of 0.1\%. Our work 
differs since it does not rely on phishing emails which are sometimes hard to identify.
An anti-phishing browser extension (AntiPhish) was given in \cite{protectphish}.
It kept track of sensitive information and warned the user whenever the user tried to
enter sensitive information into untrusted websites. Our classifier 
can be easily integrated with AntiPhish. However, AntiPhish compromised user privacy 
by keeping state of sensitive data. Other anti-phishing proposals relied on trusted 
devices, such as a user's cell phone in \cite{phoolproof}. In \cite{beyond-blacklists}, 
the authors tackled the problem of detecting malicious
websites by only analyzing their URLs using machine-learning
statistical techniques on the lexical and host-based
features of their URLs. The proposed solution achieved a prediction
accuracy around 95\%. Other studies measured the extent of typosquatting and suggested
mitigation techniques. Wang, et al.~\cite{Wang} showed that many 
typosquatting domains were active and parked with a few parking services, which served
ads on them. Similarly, \cite{jaal} showed that, for nearly 57\% of original URLs considered,
over 35\% of all possible URL variations existed on the Internet. Surprisingly, over
99\% of such similarly-named websites were considered phony. \cite{ads-portal-class} devised a methodology 
for identifying ads-portals and parked domains and found out that around 25\% of (two-level) 
.com and .net  domains were ads-portals and around 40\% of those were typosquatting. 
McAfee also studied the typosquatting problem in~\cite{mcafeestudy}. A set of 1.9 million
single-error typos was generated and  $127,381$ suspected typosquatting domains
were discovered. Alarmingly, the study also found that typosquatters targeted children-oriented domains.
Finally, McAfee added to its extension site advisor~\cite{siteadvisor}  some
capabilities for identifying typosquatters. In \cite{nwjournal-phish-1}, the authors proposed a technique to counter-against phishing and pharming attacks that is based on mutual authentication which can be easily adopted in the current systems. In \cite{nwjournal-phish-2}, the authors proposed an anomaly detection technique that is based on hidden Markov models which is very suitable to Windows environment.

To the best of our knowledge, no prior work has explored the web-fraud problem 
in the context of HTTPS and proposed analyzing server-side SSL certificates in more detail.
Our work yields a detailed analysis of SSL certificates from different domain families and a 
classifier that detects web-fraud domains based on their certificates.

Finally, some usable security studies have attempted to measure effectiveness of available anti-phishing tools and security warnings.
User studies analyzing the effectiveness of browser warning messages indicated that an
overwhelming percentage (up to 70-80\% in some cases) of users ignored them. This
observation---confirmed by recent research results (e.g., \cite{ssl-usability}
and \cite{been-warned-ssl})--- might explain the unexpected high percentage of expired and
self signed certificates that we found in the results  of all our data sets.
Furthermore, \cite{zhang2007phinding} evaluated available anti-phishing tools and points out
that only 1 out of 10 popular anti-phishing tool identified more than $90\%$ of phishing URLs correctly.
Also, \cite{wu2006security} pointed out that users failed to pay attention to the toolbar or explained away
tool's warning if the content of web pages looked legitimate.
Similarly, \cite{whalen2005gathering} highlighted, through eyetracker data, that users commonly
looked at lock icons, but rarely used the certificate information. As a result, one may wonder about the incentive for phishers and typosquatters
to utilize SSL certificates. We believe that our classifier can be used together with available tools, e.g., AntiPhish~\cite{protectphish},
which keeps track of sensitive information and warns the user whenever users enter sensitive information into insecure websites. 
In this case, phishers need to provide SSL certificates to bypass the AntiPhish block.

\section{Conclusion}\label{sec:conclusion}
In this paper, we study the prevalence of HTTPS in popular and 
legitimate domains as well as in the 
context of web-fraud, i.e., phishing and typosquatting.
To the best of our knowledge, this is the first effort to analyze 
information in SSL certificates to profile domains and 
assess their degree of trustworthiness. 
We design and build a machine-learning-based 
classifier that identifies fraudulent domains that utilize HTTPS. The classifier solely relies on 
SSL certificates of such domains, thus preserving user privacy.
Our work can be integrated with existing detection techniques to 
improve their effectiveness.
Finally, we believe that our results may serve as a motivation
to increase the adoption of HTTPS. We believe that 
aside from its intended benefits of confidentiality and authenticity, 
HTTPS can help identify web-fraud domains.


\small

\begin{thebibliography}{10}

\bibitem{alexa-ranking}
Alexa.com.
\newblock {Alexa Ranking Methodology}.
\newblock \url{http://www.alexa.com/help/traffic_learn_more}.

\bibitem{alexa}
{Alexa.com}.
\newblock {The Web Information Company}.
\newblock \url{http://www.alexa.com}.

\bibitem{spamscatter}
D.~S. Anderson, C.~Fleizach, S.~Savage, and G.~M. Voelker.
\newblock {Spamscatter: Characterizing Internet Scam Hosting Infrastructure}.
\newblock In {\em USENIX Security}, pages 1--14, 2007.

\bibitem{april07}
{Anti Phishing Working Group}.
\newblock {Phishing Activity Trends Report -- April 2007}.
\newblock \url{http://www.antiphishing.org/reports/apwg_report_april_2007.pdf},
  2007.

\bibitem{q3}
{Anti Phishing Working Group}.
\newblock {Phishing Activity Trends Report -- 3rd Quarter 2009}.
\newblock \url{http://www.antiphishing.org/reports/apwg_report_Q3_2009.pdf}, 2010.

\bibitem{jaal}
A.~Banerjee, D.~Barman, M.~Faloutsos, and L.~N. Bhuyan.
\newblock {Cyber-Fraud is One Typo Away}.
\newblock In {\em Infocom Mini-Conference}, 2008.

\bibitem{bagging}
L.~Breiman.
\newblock Bagging predictors.
\newblock {\em Machine Learning}, 24(2):123--140, 1996.

\bibitem{randforest}
L.~Breiman.
\newblock {Random Forests}.
\newblock {\em Machine Learning}, 45(1):5--32, 2001.

\bibitem{x509rfc}
{D. Cooper, et al.}
\newblock {Internet X.509 Public Key Infrastructure Certificate and CRL Profile
  (IETF RFC5280)}.
\newblock \url{http://www.ietf.org/rfc/rfc5280.txt}.

\bibitem{wwwsize}
M.~{de Kunder}.
\newblock {World Wide Web Size}.
\newblock \url{http://www.worldwidewebsize.com/}, 2010.

\bibitem{mcafee}
B.~Edelman.
\newblock {Typosquatting: Unintended Adventures In Browsing. McAfee Security Journal}, 2008.

\bibitem{been-warned-ssl}
S.~Egelman, L.~F. Cranor, and J.~Hong.
\newblock You've been warned: an empirical study of the effectiveness of web
  browser phishing warnings.
\newblock In {\em CHI}, pages 1065--1074, 2008.

\bibitem{httpseverywhere}
{Electronic Frontier Foundation}.
\newblock {HTTPS Everywhere}.
\newblock \url{https://www.eff.org/https-everywhere}, 2010.

\bibitem{googkle}
{F-Secure}.
\newblock Googkle.com installed malware by exploiting browser vulnerabilities.
\newblock \url{http://www.f-secure.com/v-descs/googkle.shtml}, 2009.

\bibitem{learnphishemails}
I.~Fette, N.~Sadeh, and A.~Tomasic.
\newblock {Learning to Detect Phishing Emails}.
\newblock In {\em WWW}, 2007.

\bibitem{boosting}
Y.~Freund and R.~Schapire.
\newblock {A decision-theoretic generalization of on-line learning and an
  application to boosting}.
\newblock In {\em CLT}, 1995.

\bibitem{parked}
{Google.com}.
\newblock {Parked Domain site}.
\newblock
  \url{http://adwords.google.com/support/aw/bin/answer.py?hl=en&answer=50002}.

\bibitem{GoogleSpellChecker}
{Google.com}.
\newblock {The Google Spell Checker}.
\newblock \url{http://www.google.co.uk/help/features.html}.

\bibitem{ssl-crypt-strength}
{H. Lee, T. Malkin and E. Nahum}.
\newblock {Cryptographic Strength of SSL/TLS Servers: Current and Recent
  Practices}.
\newblock In {\em IMC}, 2007.

\bibitem{mcafeestudy}
S.~Keats.
\newblock {What's In A Name: The State of Typo-Squatting 2007}.
\newblock \url{http://www.siteadvisor.com/studies/typo_squatters_nov2007.html},
  2007.

\bibitem{protectphish}
E.~Kirda and C.~Kruegel.
\newblock {Protecting Users Against Phishing Attacks}.
\newblock In {\em Oxford University Press 2005}.

\bibitem{ads-portal-class}
{M. Almishari and X. Yang}.
\newblock {Text-based Ads-portal Domains: Identification and Measurements}.
\newblock {\em ACM Transactions on the Web}, 4(2), 2010.

\bibitem{beyond-blacklists}
J.~Ma, L.~K. Saul, S.~Savage, and G.~M. Voelker.
\newblock {Beyond blacklists: learning to detect malicious web sites from
  suspicious URLs}.
\newblock In {\em KDD}, pages 1245--1254, 2009.

\bibitem{nwjournal-phish-1}
A.~S. Martino and X.~Perramon.
\newblock {Phishing Secrets: History, Effects, and Countermeasures}.
\newblock In {\em International Journal of Network Security}, 2010.

\bibitem{siteadvisor}
{McAfee}.
\newblock {McAfee SiteAdvisor}.
\newblock \url{http://www.siteadvisor.com/}.

\bibitem{phishloss2007}
T.~McCall.
\newblock {Gartner Survey Shows Phishing Attacks Escalated in 2007; More than
  \$3 Billion Lost to These Attacks}.
\newblock \url{http://www.gartner.com/it/page.jsp?id=565125}, 2007.

\bibitem{microsoft}
{Microsoft Researc/}.
\newblock {Screenshots of questionable advertisements}.
\newblock \url{http://research.microsoft.com/Typo-Patrol/screenshots.htm},
  2006.

\bibitem{mlearn1}
T.~Mitchell.
\newblock {\em Machine Learning}.
\newblock McGraw Hill, 1997.

\bibitem{ssl-survey-2000}
E.~Murray.
\newblock {SSL server security survey}.
\newblock
  \url{http://www.megasecurity.org/Info/ssl_servers.html}.

\bibitem{netcraft-survey}
Netcraft.
\newblock {Netcraft SSL survey}.
\newblock \url{http://news.netcraft.com/SSL-Survey}, 2008.

\bibitem{openSSL}
{OpenSSL}.
\newblock {The OpenSSL Project}.
\newblock \url{http://www.openssl.org/}.

\bibitem{phoolproof}
B.~Parno, C.~Kuo, and A.~Perrig.
\newblock {Phoolproof Phishing Prevention}.
\newblock In {\em FC}, 2006.

\bibitem{phishtank}
{Phishtank.com}.
\newblock \url{http://www.phishtank.com}.

\bibitem{dtree}
J.~R. Quinlan.
\newblock {\em {c4.5: Programs for Machine Learning}}.
\newblock Morgan Kaufmann, 1993.

\bibitem{boost-bag-c4.5}
J.~R. Quinlan.
\newblock Bagging, boosting, and C4.5.
\newblock In {\em AAAI}, 1996.

\bibitem{symmantec-blog}
Z.~Raza.
\newblock {Phishing Toolkit Attacks are Abusing SSL Certificates}.
\newblock
  \url{http://bit.ly/ACJzel}, 2009.

\bibitem{ssl-priv-public}
{S. Yilek, E. Rescorla, H. Shacham, B. Enrigh and S. Savage}.
\newblock When private keys are public: Results from the 2008 debian openssl
  vulnerability.
\newblock In {\em IMC}, 2009.

\bibitem{incr}
D.~M. Sena.
\newblock {Symantec Internet Security Threat Report Finds Malicious Activity
  Continues to Grow at a Record Pace}.
\newblock
  \url{http://www.symantec.com/about/news/release/article.jsp?prid=20090413_01}, 2009.

\bibitem{stevens2007chosen}
M.~Stevens, A.~Lenstra, and B.~De~Weger.
\newblock {Chosen-prefix collisions for MD5 and colliding X. 509 certificates
  for different identities}.
\newblock In {\em Eurocrypt}, pages 1--22, 2007.

\bibitem{stevens2009short}
M.~Stevens, A.~Sotirov, J.~Appelbaum, A.~Lenstra, D.~Molnar, D.~Osvik, and
  B.~de~Weger.
\newblock {Short chosen-prefix collisions for MD5 and the creation of a rogue
  CA certificate}.
\newblock In {\em CRYPTO}, pages 55--69, 2009.

\bibitem{Sturgeon}
W.~Sturgeon.
\newblock Serial typo-squatters target security firms.
\newblock \url{http://news.zdnet.com/2100-1009_22-5873001.html}, September
  2005.

\bibitem{ssl-usability}
J.~Sunshine, S.~Egelman, H.~Almuhimedi, N.~Atri, and L.~F. Cranor.
\newblock In {\em Usenix Security}, 2009.

\bibitem{nwjournal-phish-2}
X.~Tang, C.~N. Manikopoulos, and S.~G. Ziavras.
\newblock {Generalized Anomaly Detection Model for Windows-based Malicious
  Program Behavior}.
\newblock In {\em International Journal of Network Security}, 2008.

\bibitem{verisignurl}
{VeriSign}.
\newblock \url{http://www.verisign.com}.

\bibitem{NDSS}
Y.~Wang, D.~Beck, X.~Jiang, R.~Roussev, C.~Verbowski, S.~Chen, and S.~King.
\newblock {Automated Web Patrol with Strider HoneyMonkeys}.
\newblock In {\em NDSS}, pages 35--49, 2006.

\bibitem{Weaver}
N.~Weaver.
\newblock {HTTP is Hazardous to Your Health}.
\newblock
  \url{http://nweaver.blogspot.com/2008/05/http-is-hazardous-to-your-health.ht%
ml}, 2008.

\bibitem{Lauren}
L.~Weinstein.
\newblock {http: Must Die!}
\newblock \url{http://lauren.vortex.com/archive/000338.html}.

\bibitem{whalen2005gathering}
T.~Whalen and K.~Inkpen.
\newblock {Gathering evidence: use of visual security cues in web browsers}.
\newblock In {\em Graphics Interface}, pages 137--144, 2005.

\bibitem{wu2006security}
M.~Wu, R.~Miller, and S.~Garfinkel.
\newblock {Do security toolbars actually prevent phishing attacks?}
\newblock In {\em SIGCHI}, pages 601--610, 2006.

\bibitem{wu2006}
M.~Wu, R.~Miller, and G.~Little.
\newblock {Web wallet: preventing phishing attacks by revealing user
  intentions}.
\newblock In {\em SOUPS}, pages 102--113, 2006.

\bibitem{Wang}
{Y. Wang, D. Beck, J. Wang, C. Verbowski, and B. Daniels}.
\newblock Strider typo-patrol: Discovery and analysis of systematic
  typo-squatting.
\newblock In {\em SRUTI}, pages 31--36, July 2006.

\bibitem{zhang2007phinding}
Y.~Zhang, S.~Egelman, L.~Cranor, and J.~Hong.
\newblock {Phinding phish: Evaluating anti-phishing tools}.
\newblock In {\em NDSS}, 2007.

\end{thebibliography}

\end{document}